\documentclass{aa}

\usepackage[dvips]{graphicx}
\usepackage{txfonts}
\usepackage{natbib}
\usepackage{subfigure}

\bibpunct{(}{)}{;}{a}{}{,} 

\begin{document}


\title{The faint-end of the galaxy luminosity function in groups}

\author{R. E. Gonz\'alez \inst{1} \and %
        M. Lares \inst{2}         \and %
        D. G. Lambas \inst{2}     \and %
        C. Valotto \inst{2}}           %

\institute{%
   Departamento de Astronom\'{\i}a y Astrof\'{\i}sica,
   Pontificia Universidad Cat\'olica de Chile, Casilla 306,
   Santiago 22, Chile. 
   \and 
   Grupo de Investigaciones en Astronom\'{\i}a Te\'{o}rica y Experimental
   (IATE), 
   Observatorio Astron\'omico de C\'{o}rdoba, UNC,\\  
   and Consejo Nacional de Investigaciones Cient\'{\i}ficas y Tecnol\'ogicas.
   (CONICET), Argentina.
}

\date{Received ? ? 2005 / Accepted ? ? 2005}

\abstract{
   We compute the galaxy luminosity function in spectroscopically
   selected nearby groups and clusters. 
   Our sample comprises 728 systems
   extracted from the third release of the Sloan Digital Sky Survey in
   the redshift range $0.03 < z < 0.06$ with virial mass range
   $10^{11}M_\odot < M_{vir} < 2\times 10^{14}M_\odot$.  
   In order to
   compute the galaxy luminosity function, we apply a statistical
   background subtraction method following usually adopted techniques.
   In the $r$ band, the composite galaxy luminosity function shows a
   slope $\alpha=-1.3$ in the bright--end, and an upturn of the slope in
   the faint--end, $M_r\ga -18+5log(h)$,  to slopes $-1.9<\alpha<-1.6$.
   We find that this feature is present also in the $i,g$
   and $z$ bands, and for all explored group subsamples, irrespective of
   the group mass, number of members, integrated color or the presence
   of a hot intra-cluster gas associated to X-ray emission.

}

\maketitle
\citeindextrue


\section{Introduction}

The luminosity function (LF) of galaxies is one of the fundamental statistical
tools for describing global properties of galaxy populations. This
is important since variations in the LF in different environments can
provide clues to the different evolutionary processes of galaxies.
The LF of galaxies within groups and clusters of galaxies is then a
key property for understanding the role of the environment on galaxy
formation and evolution.  Differences in the field and cluster galaxy
LF, or between groups with different properties, could provide
statistical indicators of these environmental effects.

The field and cluster galaxy LF have been calculated with a very good
accuracy at the bright--end (brighter than $M_r \sim -17$) in different
spectroscopic surveys
\citep{blanton,mad,norberg,delapparent,depropis}.  The study of faint
galaxies ($M_r > -17$) is always restricted to small samples given the difficulties
of obtaining redshifts for large number of such objects.  
Differences between the field and cluster
galaxy LF at the bright--end may be caused by the interaction of
galaxies in different environments.  Ram pressure stripping can
inhibit star formation by exhausting the gas present in galaxies that
move fast in the intergalactic medium of rich clusters.  Similarly,
galaxy harassment can produce significant changes in the star
formation rate of galaxies.  These effects are not expected to be
important in poor clusters or groups, where the velocity dispersion is
lower, and instead, effects such as mergers or tidal interactions can be
dominant in these environments.  Studies using semi-analytic galaxies
from numerical simulations show a strong dependence of the LF on
environment \citep{rob}.  On the observational side, \citet{croton}
find a significant dependence of the LF on local galaxy density in the
2dF galaxy redshift survey.  Similar results were derived by
\citet{gar} showing a correlation between the galaxy LF and cluster
density.

A straightforward determination of the galaxy LF faint-end in galaxy
systems requires long observing times to obtain spectra of faint
objects in homogeneous samples.  Individual analysis of nearby
clusters or of the Local Group of galaxies have been performed
\citep{deady,mateo,zucker,trentham}, with the disadvantage of a low
number statistics.
On the other hand, statistical methods that perform a background
subtraction offer the possibility of analyzing large faint galaxy
samples in clusters.  
Several previous studies have made use of this technique
\citep{goto,gar,pao,popesso,valotto2}. 
In these works it is detected an excess of dwarf galaxies in the
faint-end of the LF so that a single Schechter function cannot provide
an accurate fit to the data \citep{mad}.

The large scale distribution of matter in the local universe
can give rise to spurious clumps of galaxies when observed in
projection.  
In fact, \citet{valotto1}, using a deep mock catalogue constructed from a
numerical simulation of a hierarchical universe, showed that
a biased steep determination of
the faint-end slope may be originated by projection
effects of foreground/background structures.
These effects induce a systematic
bias that cannot be corrected by subtracting background fields.  

Such systematic effects could be overcame if clusters are identified
by X-ray detection or redshift space selection, assuring the presence
of overdense regions without significant contamination along the
line of sight.
X-ray selected clusters may provide unbiased samples suitable for
determining the galaxy LF by a statistical background subtraction method, since
the hot gas confined in the gravitational potential well, responsible
of the observed thermal bremsstrahlung emission in the X--ray band, is
a reliable indicator of massive clusters.
\citet{popesso}, based on a sample
of clusters with strong X-Ray emission \citep{popesso2} calculated the
composite galaxy LF of RASS-SDSS clusters using statistical background
subtraction methods and found a steep faint-end slope $\alpha \simeq
-2$ which strongly supports the presence of a population of faint
galaxies associated to rich clusters.

\citet{popesso} results correspond to galaxies within a dense
intra-cluster medium, so it is of interest to extend this analysis to
systems without strong X-ray emission.  
Clusters and groups of galaxies selected from spectroscopic surveys
are also expected to be free from systematic identification biases as
those present in two dimensional surveys and are ideal for this extension.

The large group and cluster sample obtained from
the SDSS-DR3 spectroscopic survey by \citet[][ hereafter MZ05]{MZ05} is
suitable to calculate a reliable composite galaxy LF in galaxy systems
of a wide range of virial masses extending previous results in rich
clusters to more common galaxy environments.  The size of this sample
allows to analyze subsamples with different group properties (e.g.,
mass, sizes, color, morphology), so that the LF
dependence on group characteristics can be assessed.

In this paper we study the galaxy LF in groups and clusters from MZ05,
using a statistical background subtraction method from SDSS photometric data.  We
derive a reliable galaxy LF down to $M_r \simeq -14$ and explore
different subsamples according to group and galaxy properties.  The
paper is structured as follows. In section 2 we describe the data used
to calculate the composite LF in clusters. In section 3 we describe
the method of statistical background subtraction adopted and the fitting
functions.  In section 4 we show the results for different subsamples
in order to test for a possible dependence of the LF with group
properties, we also compare our results with other studies in the
field and in clusters.  Finally, in section 5 we summarize our
conclusions.  Distance-dependent quantities are calculated using a
Hubble parameter $H_0=100\,h\,Km\,s^{-1}\,Mpc^{-1}$.


\section{The data \label{data}}

In this paper we use the Third Data Release of the Sloan Digital Sky
Survey \citep[SDSS DR3,][]{abaz2}, in fields centred in MZ05 groups
and clusters.  SDSS provides images and spectroscopic CCD data
at high galactic latitudes, using a dedicated wide-field 2.5m
telescope at Apache Point Observatory in South-East New Mexico. This
survey includes imaging data in five bands over $5282\,deg^2$,
within a  photometric and astrometric catalogue of 141 million
objects.  
The photometric data were also taken from the SDSS DR3. We extracted
circular shape fields centred in each group, subtending a projected
radius of $1.2 \,h^{-1}\,Mpc$ at the group euclidean distance.  The
fields include galaxies in the five photometric bands with a $95\%$
completeness at $u,g,r,i,z=22.0,22.2,22.2,21.3$ and $20.5$.

The galaxy groups used were taken from the galaxy group
sample constructed by MZ05. In order to identify groups of
galaxies in the SDSS DR3, they developed an algorithm based on the
friends-of-friends percolation method of \citet{hgpm}.
The groups have an
overdensity threshold of $\delta \rho / \rho =80$ and a line-of-sight
linking length parameter $V_0=200\;Km\;s^{-1}$. The identification
takes into account the variation of the number density of galaxies due
to the apparent magnitude limit and the sky coverage of the galaxy
catalogue.  Also, the authors apply an iterative method in dense
regions in order to turn artificial systems aside and achieve a better
determination of the group centers \citep{diaz}.  The final sample of
MZ05 comprises 10864 groups with at least four members.  We used a
fraction of these systems, confining the sample to the nearby groups
with $0.03<z<0.06$ and whose corresponding SDSS fields lack evident
incompleteness.

The redshift restriction was adopted to have a fair sample of groups
for the following reasons: Although the linking length parameters used
in the identification have been tested in mock catalogues, groups at
too low redshift ($z<0.03$) have an important contribution from
peculiar velocities, and also, fields corresponding to these groups
cover very large areas.  In order to reach the faint-end of the galaxy
LF, we adopted an upper redshift limit $z=0.06$, that corresponds
to an absolute magnitude $M_r = -14.08$ at the limiting apparent
magnitude $m_r = 22.2$.
From the total sample of MZ05 groups, the redshift restriction give us
a subsample of 1782 systems.  We have further removed groups which are
affected either by mask-related completeness problems, or by large
fluctuations in the number counts of galaxies in neighbouring regions
surrounding the group centers.  This conservative restriction
provides  728 groups which are best suited for a reliable statistical
background subtraction method and will be used hereafter.

We have also searched for X-ray counterparts of our groups in the RASS-SDSS
catalogue using a simple criterion of proximity to associate MZ05
systems to RASS X-ray clusters.  The RASS-SDSS is a compendium of
X-ray emitting clusters of galaxies based on the ROSAT All Sky Survey
(RASS) and the SDSS \citep{popesso2}.  Adopting a maximum allowed
separation between 
a MZ05 groups and a RASS-SDSS objects 
of $0.5$Mpc, all the RASS clusters within our
redshift interval are matched by one MZ05 group belonging to our
sample, resulting 15 clusters in common.

In order to characterize the environment of the groups, we used the
virial mass and the number of members to define appropriate
subsamples.  We also defined two additional parameters that may be
useful to characterize the environment of each group: the integrated
color index and the dominant galaxy estimator.

The integrated color index of each group was calculated by integrating
the luminosity of each spectroscopic member galaxy in the $u$ and $r$ bands, $L_u$
and $L_r$ respectively, and obtaining a global color of the group,
according to:

\begin{equation}
u-r = -2.5Log \bigg{[} \frac{\sum L_u}{\sum L_r} \bigg{].}
\end{equation}

\noindent

The dominant galaxy estimator is defined by the usual criterion of the
difference between the magnitudes of the brightest and the
third--ranked  member of each group, \mbox{$|M_{r1} - M_{r3}|$}.

\begin{figure}  
  \resizebox{\hsize}{!} {%
  \includegraphics[width=\columnwidth]{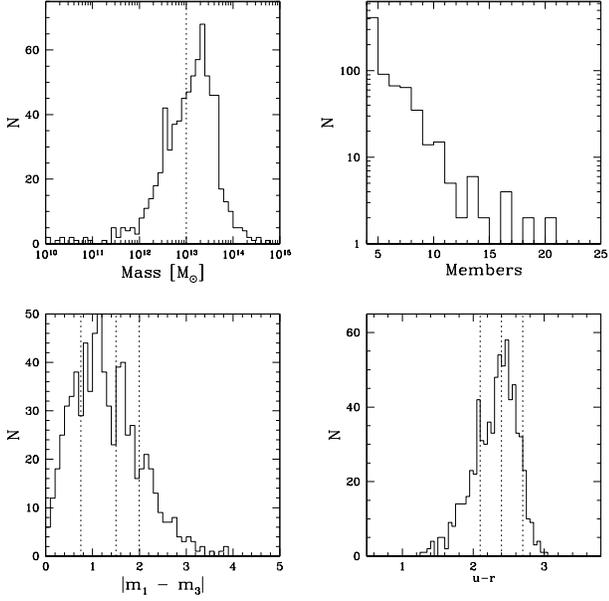}}


  \caption{%
   Main properties of the 728 groups of the sample of groups.
   Top panels: 
   Distribution of virial mass and number of spectroscopic members.
   Bottom panels: Distribution of dominant galaxy
   estimator and integrated color index.
   Dotted lines represent cuts of the sample for 
   further analysis.}

  \label{gp}
\end{figure}

Figure \ref{gp} shows the distribution of the main properties of our
total sample of 728 groups. The dashed lines correspond to the  different
cuts adopted.

\section{Background estimation and Luminosity Function computation.}

Our study of the galaxy LF is based on a statistical 
background subtraction method
applied in order to statistically remove the contribution of
foreground and background galaxies in the fields of MZ05 groups.
Local variations in the background number density of galaxies can be
taken into account by using a local rather a global background, close
enough to consider these local variations, but beyond the group
region. 
Taking into account these considerations, we derived a local
background using galaxies in a ring $2\,h^{-1}\,Mpc
<R_p<5\,h^{-1}\,Mpc$ centred in each group.
We find that the results of using either local or global statistical 
background
subtraction methods are consistent with each other within the rms
scatter of each magnitude bin.  However, the advantage of a global
background relies in the fact that it can be very stable statistically
if derived from a
sufficiently large area. Therefore we adopted this global background
to perform the statistical decontamination in the group fields.
The steps followed to decontaminate the galaxy counts are:
1- We selected $4816$ random fields within the area covered by the
spectroscopic SDSS DR3. Each field has an angular radius $8\,arcmin$
so that the total number of fields comprise $\sim3.6\times10^6$
galaxies in an area of $268.97$ square degrees.  
2- From these fields, we constructed a distribution of apparent
magnitudes binned in 0.25 mag intervals.  
3- For each group we subtracted this background distribution
normalized to the relative area of the group region and the
background. This procedure was applied to three  regions
centred in each group defined by $R_p < 0.3$ $h^{-1}$ Mpc, 
$R_p < 0.5$ $h^{-1}$ Mpc, 
$R_p < 0.7$ $h^{-1}$ Mpc and $0.3 < R_p < 0.7$ $h^{-1}$ Mpc.
4- We summed the excess in the number of galaxies for each group in each
apparent magnitude interval. We assume that the excess galaxies are at
the mean group redshift so that for each group we obtain an absolute
magnitude distribution.
5- We defined different subsamples according to group
properties, as discussed in section 4, and summed the contributions to the
galaxy LF of each group of the subsamples.
We have considered an apparent limiting magnitude $r = 22.2$
corresponding to a $95\%$ completeness.  
At the maximum redshift of our sample of groups, $z<0.06$, this
corresponds to an absolute limiting magnitude $M_r = -14.25$.  
Absolute magnitudes were computed using $M=m - 25 -5log_{10}(D_L),$
where $D_L$ is the luminosity distance in $Mpc$. 
Due to the low redshift of the sample, K-corrections are negligible.

Since it is observed an upturn in the galaxy LF derived we adopted
two different Schechter functions \citep{schechter} for the bright
and the faint region respectively (see next section for details): 
\begin{eqnarray} \qquad \qquad \phi
(M)dM = 0{.}4 \,ln(10)\, \phi^* \,e^{-X} \,X^{\alpha + 1}\,dM
\nonumber \end{eqnarray} \noindent where $ X = 10^{0{.}4(M^* - M)}$,
the parameter $M^*$ refers to the characteristic luminosity;
$\alpha$ is the faint-end slope indicating the
relative importance of a population of low luminosity galaxies, and
$\phi^*$ is the LF normalization.
Using a maximum likelihood estimator we fit the best Schechter
functions in both the bright and in faint region of the LF. 
In tables 1 and 2 we show fits for different subsamples of our data. 
Errors on these tables correspond to 1$\sigma$ contours (68\%
confidence), following Poisson statistics in the galaxy counts.


\section{Analysis and results}

\begin{figure}  
\resizebox{\hsize}{!}{\includegraphics[width=\columnwidth]{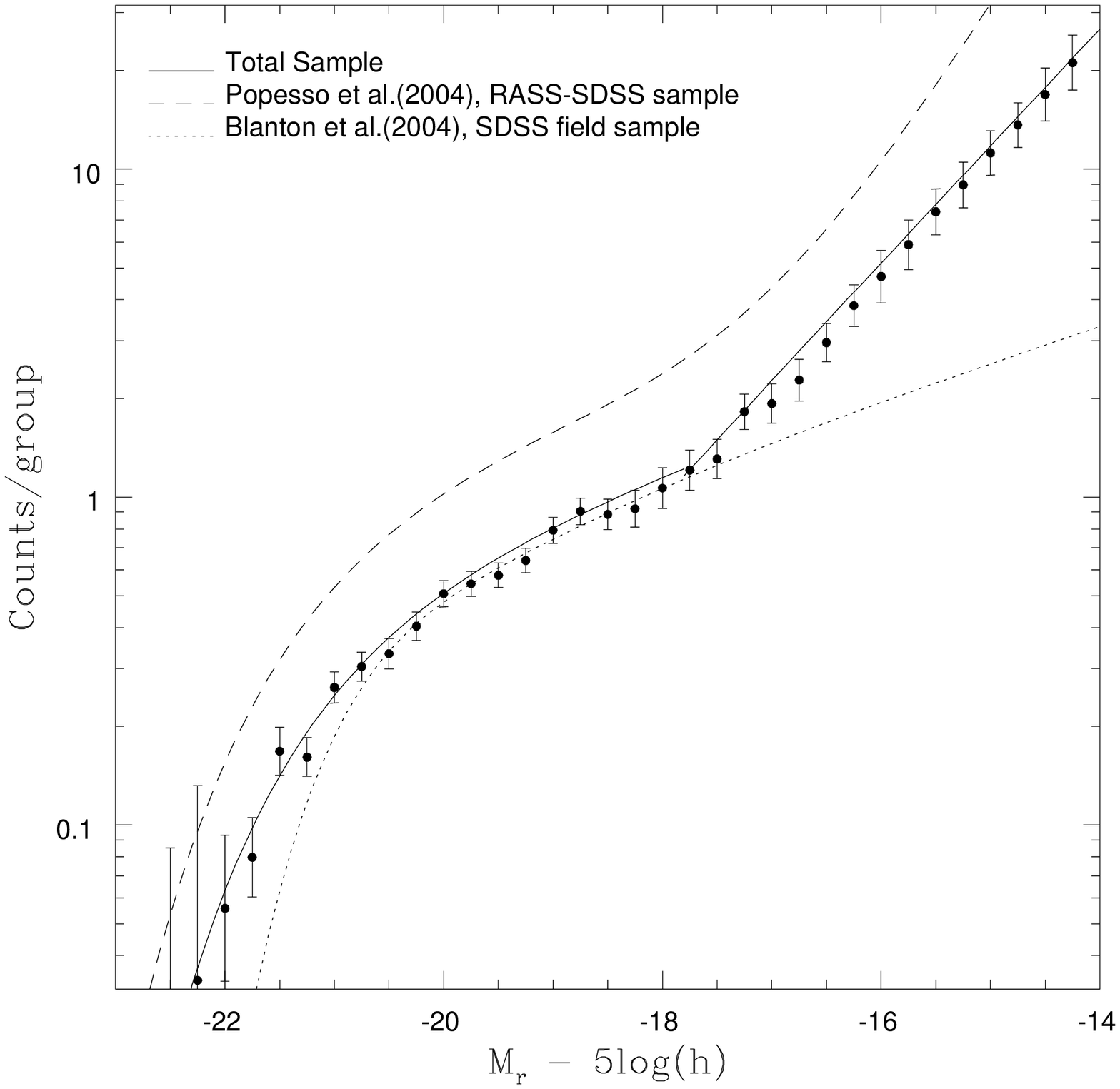}}
  \caption{
   r-band composite galaxy LF for the total sample of groups calculated
   within $0.5 h^{-1}$ Mpc from group centres.
   The solid line corresponds to the best two Schechter function fits
   obtained from a maximum likelyhood estimator with an upturn limit at
   $M_r = -18$ (see \ref{table2} for parameter values).
   For comparison  we show with an arbitrary normalization \cite{popesso}
   and \cite{blanton} LF determinations of galaxies in X-ray
   clusters and in the field, respectively.}
  \label{lfcomp}
\end{figure}

We have typically $\simeq 8000\,-\,12000$ galaxies per square degree
in each group field, brighter than $m_r=22.2$, out of which 
$\simeq 100\,-\,4000$ are in
excess within $0.7\;h^{-1}\; Mpc$ with respect to the mean global
background.
This allows us to construct the distribution of absolute magnitudes, and
use it to compute the galaxy LF.
The results for the composite LF of our total sample of 728 groups is shown in figure
\ref{lfcomp}, the faintest
bin ($-14 < M_r < -14.25$) has a Poisson uncertainty less than
$10\%$.
In this figure we also show, with arbitrary normalization,
\citet{popesso} results for galaxies in 97 clusters of the RASS-SDSS
sample, and the field LF determination of \citet{blanton}.  
We can appreciate that the bright-ends are similar, and that X-ray
clusters and optical groups also have comparable steep
faint-ends, with the LF of galaxies in groups slightly flatter than
that of galaxies in the X--ray selected sample.  
We notice that although MZ05 groups and X-ray clusters in
\citet{popesso} may differ considerably in their properties, the
galaxy LF are remarkably similar, indicating that a rising LF at the
faint end is a generic feature of galaxy systems.  
A single Schechter function does not provide an accurate fit to the
observed galaxy LF due to this upturn occurring approximately at
$M_r=-18$.  
Two separate Schechter functions provide an appropriate fit to
both the bright and faint regions as can be appreciated in fig \ref{lfcomp}.  
By comparison of different determinations of the galaxy LF, it can be
seen the presence of a statistically significant excess of extremely
faint galaxies ($M_r>-17.$) in our analysis and in
\citet{popesso}, with respect to the field galaxy LF obtained by
\citet{blanton}. 
We notice, however, the lack of faint galaxies in the spectroscopic surveys 
so that the faint-end slope in the field can be strongly affected by completeness corrections.
The agreement of our composite LF with \citet{blanton} determination
at intermediate magnitudes and in the bright-end where there is a
reasonable completeness in \cite{blanton} provides confidence in our
statistical background subtraction procedure.  
%
%
We have assessed the reliability of the faint-end slope determination taking into account the Poisson 
uncertainty of the global background fluctuation $\sim 1\%$ and the fact that we achive $\sim 95 \%$ completeness in the faintest magnitude bins. It would be required a relative error as large as $40\%$ in order to make our faint end determination consistent with \cite{blanton}.

\begin{figure*}
\resizebox{\hsize}{!}{\includegraphics[width=\columnwidth]{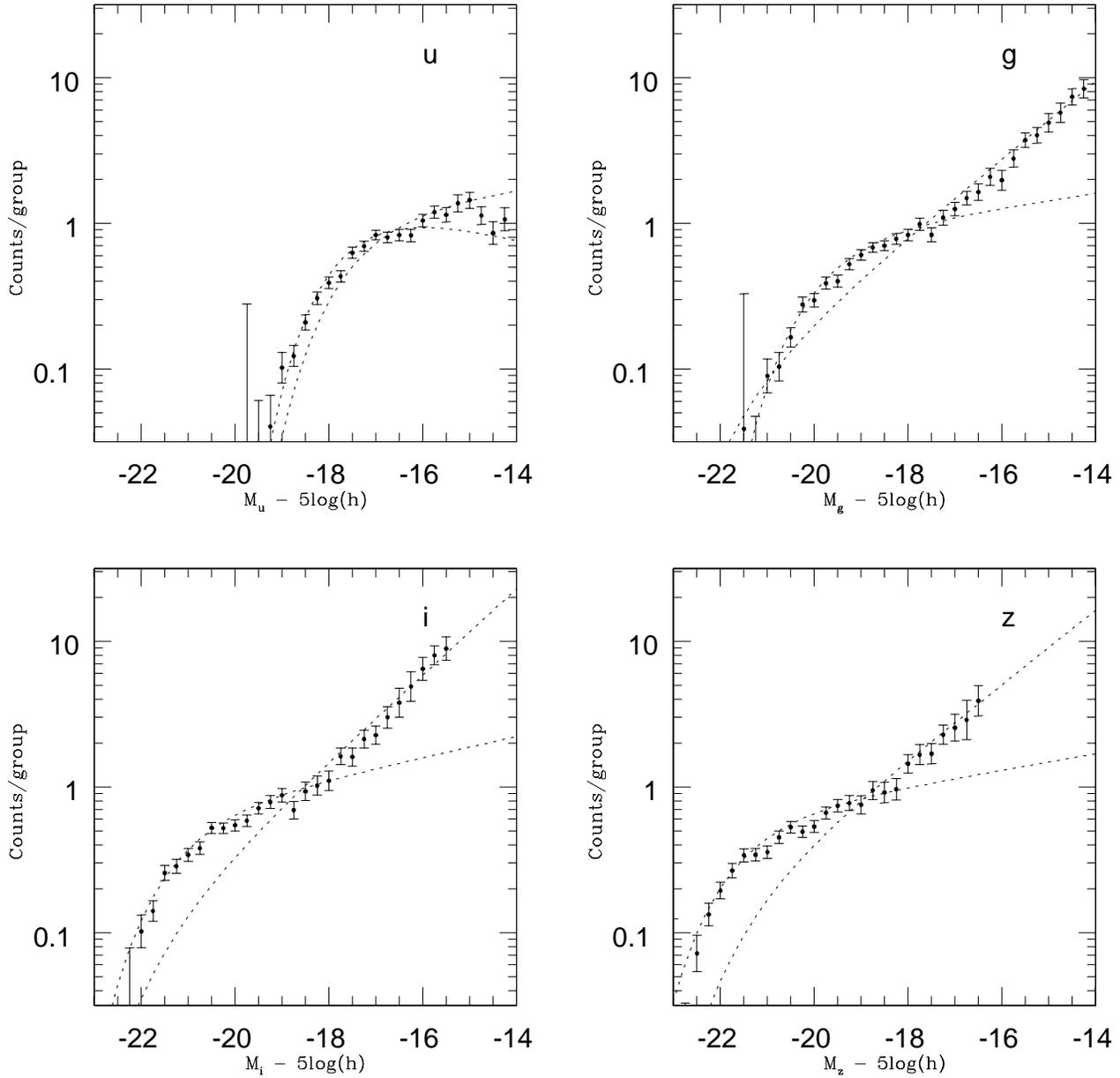}}
  \caption{Composite galaxy LF in the $u,g,i,z$ photometric bands 
   calculated for the total group sample within $0.5\,h^{-1}\, Mpc$ 
   from group centres.}
  \label{5bandslf}
\end{figure*}

\begin{figure}  
  \resizebox{\hsize}{!} {%
  \includegraphics[width=\columnwidth]{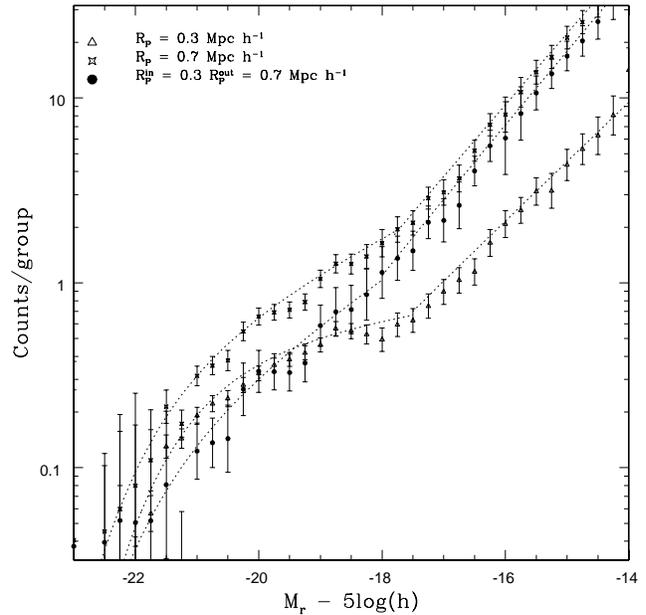}}
  \caption{%
   Composite r-band galaxy LF for different group centric distance ranges.}
  \label{gaper}
\end{figure}

We have also derived the composite galaxy LF in the other SDSS
photometric bands ($u$,$g$,$i$,$z$).  
The results are shown in figure \ref{5bandslf}, and in table 1 we list
the two component Schechter function fits to each LF, with upturn
limits in $M_g, M_i, M_z = -17.5, -18.5$ and $-19$.  
Therefore, the galaxy LF requires two Schechter functions to provide a
suitable fit to the observations, except for the $u$-band where we can
only see the bright-end. 
For this band, we would expect the upturn 
close to the magnitude limit, since the mean color $u-r \simeq 2.4$ of
groups would imply an upturn at $M_u \simeq -15.6$.
Photometric errors are small in the $g$,$r$ and $i$ band, where the
upturn of the LF is clear; in $u$ and $z$-bands photometric errors are
greater and the completeness limit is $\simeq 0.5$ magnitudes
brighter, although the upturn in the $z$-band is clear.  
This is consistent with a constant extent of the bright region of $\sim
4.5$ magnitudes in all bands.
In order to explore a possible dependence of the dwarf galaxy
population on group-centric distance, we consider the counts of
galaxies for each galaxy group in four regions,
$0<R_p<0.3\;h^{-1}\;Mpc$,
$\;0<R_p<0.7\;h^{-1}\;Mpc$,$\;0.3<R_p<0.7\;h^{-1}\;Mpc$.
These results are shown in figure \ref{gaper} and in table 2 where it
can be appreciated that the faint--end slope is steeper in the
outskirts of clusters, where a larger fraction of star-forming
galaxies is expected by the morphological segregation of galaxies in
clusters \citep{andreon,domin}.  
Although with a lower statistical significance, bright galaxies are
found to be more strongly concentrated.


\subsection{Galaxy LF dependence on group properties}

\begin{figure}  
  \resizebox{\hsize}{!} {%
  \includegraphics[width=\columnwidth]{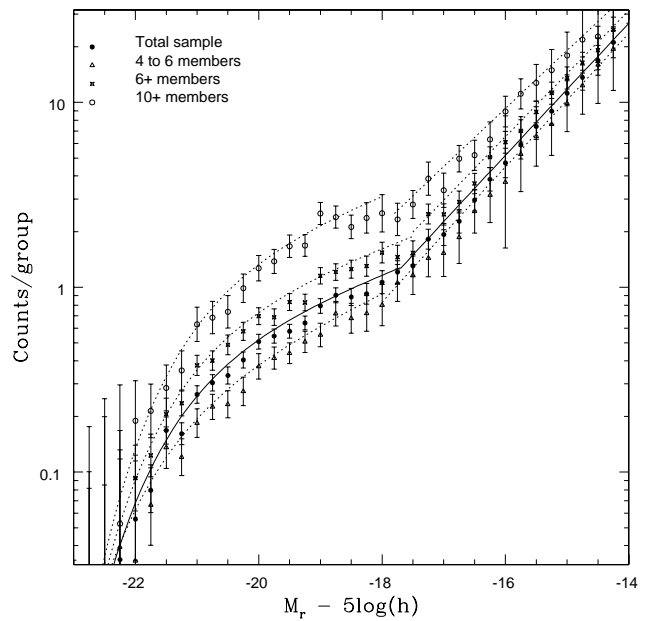}}
  \caption{Dependence of the composite galaxy LF, on the number of
           group members.
           It can be appreciated that the faint-end 
           slope is slightly flatter for groups with more members.}
  \label{gmembers}
\end{figure}

\begin{figure}
  \resizebox{\hsize}{!} {%
  \includegraphics[width=\columnwidth]{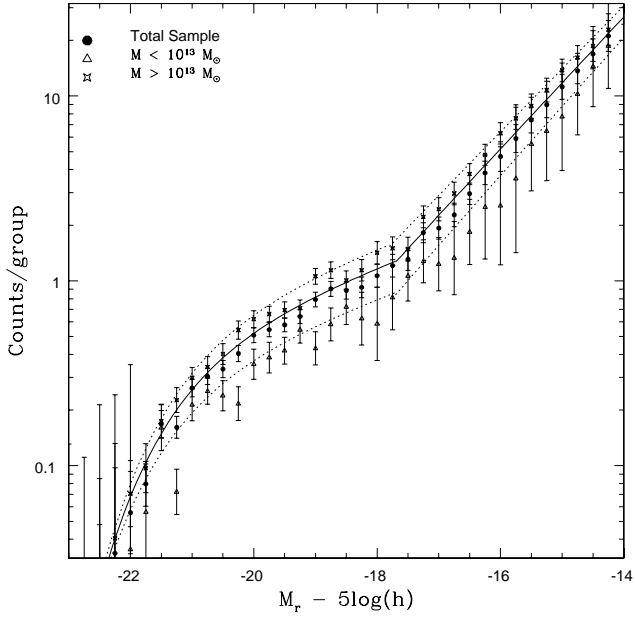}}
  \caption{Composite galaxy LF for different group mass ranges.}
  \label{gmass}
\end{figure}

We have explored several subsamples of groups according to the number
of members, and total virial mass.  
We have also considered two additional parameters described in
section \ref{data}, the group integrated color
index and the dominant galaxy contrast
parameter.
By studying the galaxy LF in subsamples according to different ranges in the number of
members, we conclude that there are no important differences in the
results although we detect that in the richer systems the galaxy LF has a 
slightly flatter
faint-end slope as we can see in figure \ref{gmembers}.  
The results of the composite LF for groups of different virial mass
can be appreciated in figure \ref{gmass} and table 2 where it can be seen 
that there are
no substantial differences between high and low group
mass, although here again there is a trend for massive groups to have a
flatter LF faint-end slope.  
This result is consistent with the fact that the galaxy LF in groups
and in rich, X-ray emitting clusters are similar.

\begin{figure}
  \resizebox{\hsize}{!} {%
  \includegraphics[width=\columnwidth]{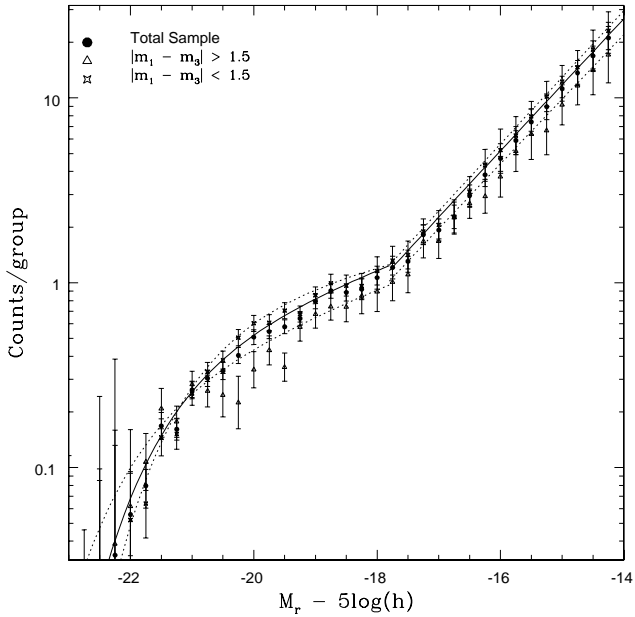}}
  \caption{Composite galaxy LF for groups with
           different dominant galaxy estimator.}
  \label{gdom}
\end{figure}

\begin{figure}
  \resizebox{\hsize}{!} {%
  \includegraphics[width=\columnwidth]{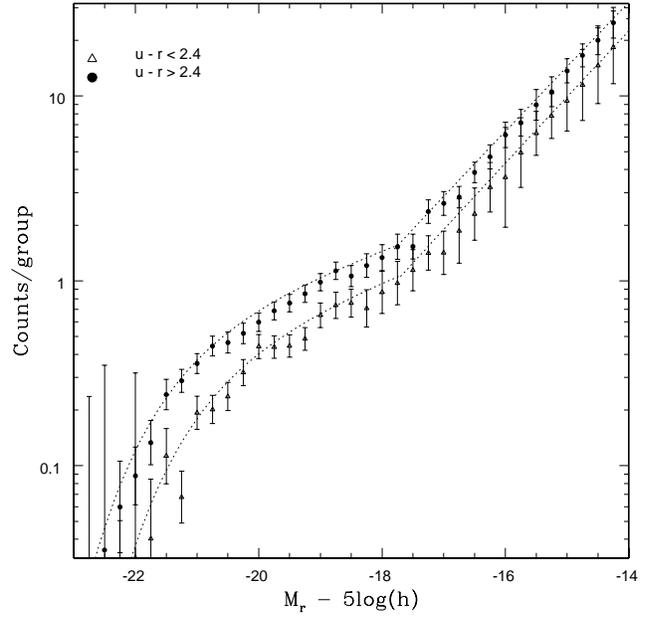}}
  \caption{Composite galaxy LF for groups with different ranges of 
           integrated color index.
  \label{gcolor}}
\end{figure}

In order to explore the presence of possible evolution effects, we have divided
the group sample according to both the group integrated color index
and the dominant galaxy estimator using spectroscopic menbers, 
without imposing no further
restrictions on the other group parameters.  
By comparing the results for subsamples defined by extreme values of
these parameters, we aim to find possible differences in the galaxy LF
as a result of evolution in groups.  
According to our analysis, no relevant differences are obtained.
By inspection to figures \ref{gcolor} and \ref{gdom}, and table 2, it
can be appreciated the lack of significant differences in the global
shape of the galaxy luminosity function regardless of the parameter of the subsample
analyzed.  
The results indicate that irrespective of the integrated color index,
or the presence of a dominant galaxy, groups have a similar
galaxy LF.  
Our findings support the existence of a similar large fraction of
dwarf galaxies, regardless of mass, average color, or
dominant galaxy luminosity contrast of the groups.

\begin{figure}
  \includegraphics[width=\columnwidth]{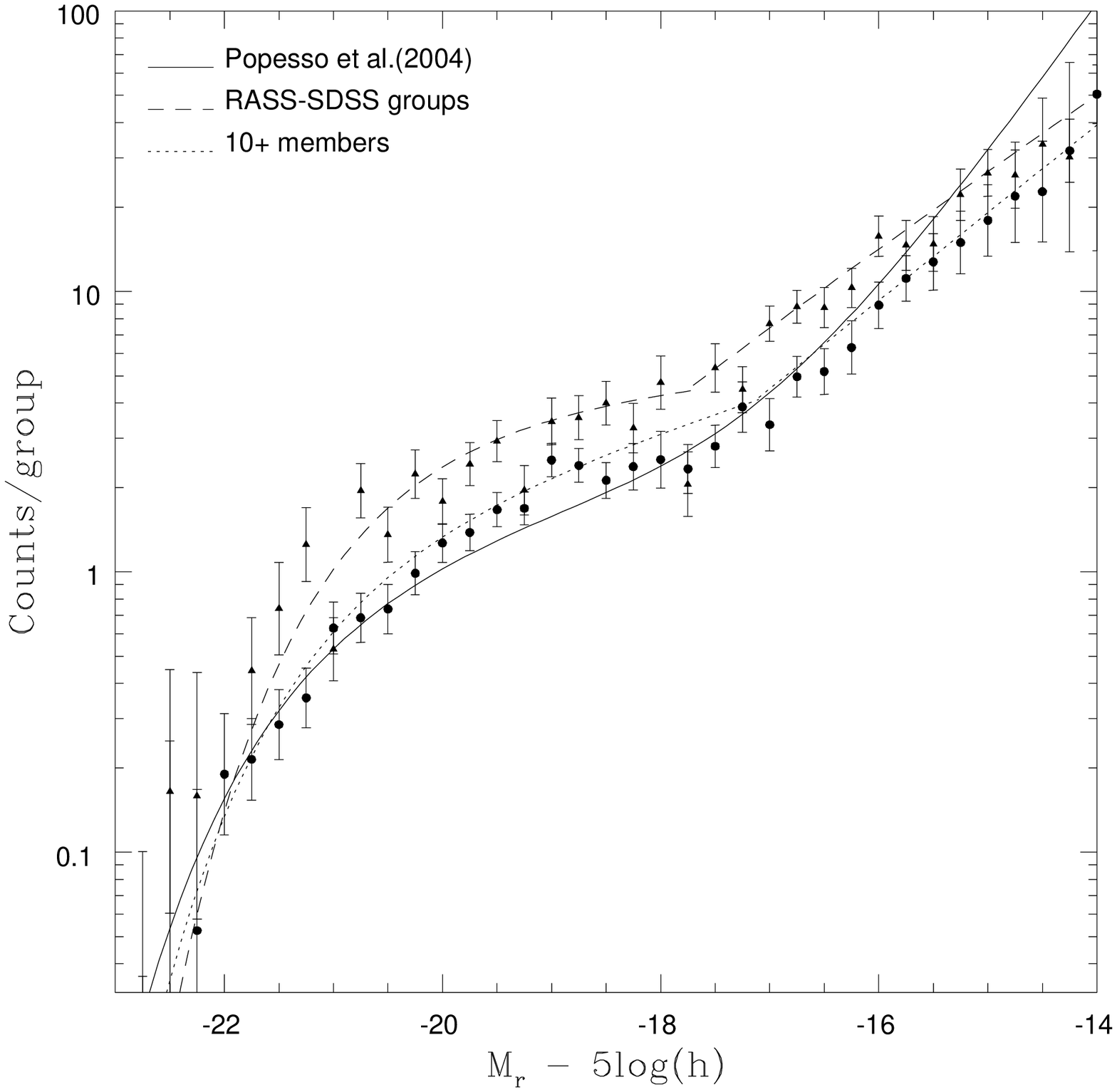}

  \caption{
   r-band composite galaxy LF of the 15 RASS-SDSS clusters in our sample.
   For comparison we plot the composite galaxy LF of the groups with 
   more than 10 members, and 
   with an arbitrary normalization, the galaxy LF
   of the full RASS-SDSS sample determined by \cite{popesso}.}
   \label{grass} 
\end{figure}

Finally, we plot in \ref{grass} the composite galaxy LF for the
subsample of 15 X-ray emitting systems described in section 2.
It can be seen in this figure that irrespective of the presence of a
hot intra-group gas, the galaxy LF faint-end slopes of rich groups are
similar.
By inspection to table 2 it can be appreciated that the faint-end
slope $\alpha= -1.68 \pm 0.10$ for the 15 groups coincident with the
RASS-SDSS sample is consistent with $\alpha = -1.78 \pm 0.06$ derived
for rich groups with more than 10 members.


\begin{table}
\begin{center}
\label{table1}
\begin{tabular}{cc@{$\pm$}cc@{$\pm$}cc@{$\pm$}cc@{$\pm$}cc}
\hline
band &
\multicolumn{2}{c} {$\alpha_{bright}$}           &
\multicolumn{2}{c} {$M^{\ast}_{bright}$}         &
\multicolumn{2}{c} {$\alpha_{faint}$}            & 
\multicolumn{2}{c} {$M^{\ast}_{faint}$}          &
{$\phi^{\ast}_b/\phi^{\ast}_f$} \\
\hline
u&-0.79&0.06&-17.68&0.05&-1.12&0.15&-17.67 &0.4&1.3913  \\
g&-1.12&0.06&-20.10&0.10&-1.67&0.03&-21.95 &0.5&11.695 \\
r&-1.31&0.04&-21.42&0.12&-1.89&0.04&-21.94 &0.5&11.509   \\
i&-1.18&0.08&-21.50&0.17&-1.74&0.03&-21.97 &0.5&6.4023   \\
z&-1.14&0.05&-21.91&0.15&-1.64&0.10&-21.85 &0.6&3.8122    \\
\hline
\end{tabular}
\caption{Galaxy LF bright and faint-end Schechter fits parameters
         in the five photometric bands within $0.5\,h^{-1}\,Mpc$ 
         of group-centric distance.}
\end{center}
\end{table}


\begin{table*}
\begin{center}
\label{table2}
\begin{tabular}{lc@{$\pm$}cc@{$\pm$}cc@{$\pm$}cc@{$\pm$}cc}
\hline
Dataset &
\multicolumn{2}{c} {$\alpha_{bright}$}           &
\multicolumn{2}{c} {$M^{\ast}_{bright}$}         &
\multicolumn{2}{c} {$\alpha_{faint}$}            & 
\multicolumn{2}{c} {$M^{\ast}_{faint}$}          &
{$\phi^{\ast}_b/\phi^{\ast}_f$} \\
\hline
Background Subtraction \\
in total sample&-1.31&0.04&-21.40&0.12&-1.89&0.04&-21.94&0.5&11.509 \\ 	
\hline
Group centric distance [$h^{-1}$ Mpc] \\
0.3  &-1.14&0.03&-21.2&0.2&-1.80&0.03&-21.42&0.7&10.759 \\
0.7  &-1.44&0.04&-21.7&0.2&-1.96&0.03&-21.52&0.8&5.7132 \\
0.3-0.7  &-1.58&0.05&-21.76&0.25&-1.99&0.03&-20.98&0.7&1.869 \\
\hline
Members \\
4 to 6    &-1.39&0.08&-21.72&0.20&-1.90&0.08&-21.70&0.6&6.2556 \\
6 or more &-1.30&0.06&-21.35&0.15&-1.85&0.03&-21.80&0.7&9.5694 \\ 
10 or more&-1.31&0.08&-21.22&0.20&-1.78&0.06&-21.95&0.8&9.9988 \\
\hline
Mass [$M_\odot$] \\
$10^{10}-10^{13}$&-1.30&0.09&-21.52&0.25&-1.93&0.10&-21.30&0.8&7.7569 \\
$10^{13}-10^{15}$&-1.31&0.05&-21.35&0.20&-1.85&0.03&-21.77&0.6&8.2840 \\
\hline
Integrated group Color u-r\\
0.0 - 2.4&-1.32&0.08&-21.17&0.19&-1.89&0.07&-21.72&0.9&9.9962 \\
2.4 - 4.0&-1.30&0.06&-21.62&0.20&-1.87&0.04&-21.65&0.9&7.8454 \\
0.0 - 2.1&-1.16&0.13&-20.10&0.22&-1.86&0.40&-17.82&1.2&0.7694 \\
2.7 - 4.0&-1.30&0.12&-21.95&0.21&-1.83&0.13&-19.05&1.4&0.6582 \\
\hline
Dominant galaxy contrast [$M_{r1}-M_{r3}$] \\
1.5 - 5  &-1.33&0.03&-21.97&0.18&-1.87&0.06&-21.88&0.9&6.9975 \\
0 - 1.5  &-1.19&0.06&-20.98&0.13&-1.90&0.05&-21.97&1.2&18.992 \\	
0 - 0.75 &-1.03&0.09&-20.63&0.15&-1.83&0.06&-21.87&0.5&16.856 \\
2.0 - 5  &-0.51&0.12&-21.17&0.17&-1.74&0.05&-21.82&0.6&9.5708 \\
\hline
RASS-SDSS groups sample  \\
15 groups&-1.09&0.08&-20.75&0.14&-1.68&0.10&-20.57&1.4&4.4755 \\
\hline
\end{tabular}
\caption{r-band LF bright and faint regions Schechter fits for the different
         subsamples. Otherwise stated, the composite LF is determined
         within $0.5 \,h^{-1}$ Mpc.}
\end{center}
\end{table*}


\section{Conclusions}

We provide on firm statistical basis the presence of a steep faint-end
in the galaxy LF in groups and poor clusters, with small variations
according to group properties. 
The results are consistent with those obtained in X-ray clusters by
\citet{popesso}.
The spurious detections of clumps in projection is an usual
disadvantage of statistical background subtraction methods. 
We argue that this not introduce a signifcant bias when selecting the centers of the
clusters in redshift space.  
We also notice that since the groups were selected in redshift space,
they are not likely to be affected by the projection effects explored
in \citet{valotto1}.  
Our analysis is complete down to $M_r=-14.25$ and thus provide a
reliable determination of the relative fraction of faint objects in
galaxy systems.  
Overall, the shape of the galaxy LF cannot be described by a single
Schechter function, mainly due to an upturn of the faint-end slope,
occurring at $M_r \sim -18$.
The observed faint-end slope is quite steep, $\alpha \simeq -1.9$
indicating a large fraction of faint galaxies in groups and clusters.
This is consistent with \citet{popesso} results for X-ray clusters
providing strong evidence that the upturn is not an exclusive feature
of galaxies in a hot gas environment.
Consistent with previous works, we also find evidence that the
characteristic luminosity $M^{\ast}$ is brighter in groups and
clusters than in the field.

We have analyzed several subsamples of groups with different
properties such as virial mass, number of members, global color index,
and presence of a dominant galaxy.  
In all cases, we find only small variations of the shape of the LF in
the faint-end.  
This suggests a universal nature of the shape of the galaxy LF in
galaxy system.

\begin{acknowledgements}

This work was supported by the Latin American-European Network on
Astrophysics and Cosmology of the European Union's ALFA Programme, the
Consejo Nacional de Investigaciones Cient\'{\i}ficas y Tecnol\'ogicas
(CONICET), the Secretar\'{\i}a de Ciencia y T\'ecnica (UNC), and the
Agencia C\'ordoba Ciencia.

Funding for the creation and distribution of the SDSS Archive has been
provided by the Alfred P. Sloan Foundation, the Participating
Institutions, the National Aeronautics and Space Administration, the
National Science Foundation, the U.S. Department of Energy, the
Japanese Monbukagakusho, and the Max Planck Society. The SDSS Web site
is http:\/\/www.sdss.org\/.

The SDSS is managed by the Astrophysical Research Consortium (ARC) for
the Participating Institutions. The Participating Institutions are The
University of Chicago, Fermilab, the Institute for Advanced Study, the
Japan Participation Group, The Johns Hopkins University, the Korean
Scientist Group, Los Alamos National Laboratory, the
Max-Planck-Institute for Astronomy (MPIA), the Max-Planck-Institute
for Astrophysics (MPA), New Mexico State University, University of
Pittsburgh, Princeton University, the United States Naval Observatory,
and the University of Washington.

\end{acknowledgements}

\bibliographystyle{aa} 
\bibliography{references}   

\end{document}